\pdfoutput=1
\documentclass[11pt,twoside,a4paper,cmspaper,final,collab]{cms-tdr}
\def\svnVersion{b9e925b}\def\svnDate{2022/07/05}\def\cmsCernNoTag{CERN-EP-2021-254}\def\cmsCernDate{\today}\def\cmsMessage{Published in Physical Review Letters as \href{http://dx.doi.org/10.1103/PhysRevLett.129.022001}{\doi{10.1103/PhysRevLett.129.022001}.}}
\begin{document}\cmsNoteHeader{HIN-20-001}

\newcommand{\PbPb}{\ensuremath{\text{PbPb}}\xspace}
\newcommand{\vtwofour}{\ensuremath{v_{2}\{4\}}\xspace}
\newcommand{\vtwotwo}{\ensuremath{v_{2}\{2\}}\xspace}
\newcommand{\vtwo}{\ensuremath{v_{2}}\xspace}
\newcommand{\vn}{\ensuremath{v_{\mathrm{n}}}\xspace}
\newcommand{\minv}{\ensuremath{m_{\text{inv}}}\xspace}
\newcommand{\vtwosigbkg}{\ensuremath{v_{2}^{\text{sig+bkg}}\{4\}}\xspace}
\newcommand{\cnfour}{\ensuremath{c_{\mathrm{n}}\{4\}}\xspace}
\newcommand{\dnfour}{\ensuremath{d_{\mathrm{n}}\{4\}}\xspace}
\newcommand{\vnfour}{\ensuremath{v_{\mathrm{n}}\{4\}}\xspace}
\newcommand{\vtworatio}{\ensuremath{\vtwofour/\vtwotwo}\xspace}
\newcommand{\llangle}{\ensuremath{\langle\langle}}
\newcommand{\rrangle}{\ensuremath{\rangle\rangle}}

\cmsNoteHeader{HIN-20-001}
\title{Probing charm quark dynamics via multiparticle correlations in \texorpdfstring{\PbPb}{PbPb} collisions at \texorpdfstring{$\sqrtsNN=5.02\TeV$}{sqrt(sNN) = 5.02 TeV}}

\date{\today}

\abstract{
Multiparticle azimuthal correlations of prompt \PDz mesons are measured in \PbPb collisions at a nucleon-nucleon center-of-mass energy of $\sqrtsNN=5.02\TeV$. For the first time, a four-particle cumulant method is used to extract the second  Fourier coefficient of the azimuthal distribution (\vtwo) of \PDz mesons as a function of event centrality  and the \PDz transverse momentum.  The ratios of the four-particle \vtwo values to previously measured two-particle  cumulant results provide direct experimental access to event-by-event fluctuations of charm quark azimuthal anisotropies.  These ratios are also found to be comparable to those  of inclusive charged particles in the event. However, hints of deviations are seen in the most central and peripheral collisions.  To investigate the origin of flow fluctuations in the charm sector, these measurements are compared to a model implementing fluctuations of  charm quark energy loss via collisional or radiative processes in the quark-gluon  plasma. These models cannot quantitatively describe the data over the full transverse momentum and centrality ranges,  although the calculations with collisional energy loss provide a better description of the data. 
}

\hypersetup{%
pdfauthor={CMS Collaboration},%
pdftitle={Probing charm quark dynamics via multiparticle azimuthal correlations of D0 mesons in PbPb collisions at sqrt(sNN) = 5.02 TeV},%
pdfsubject={CMS},%
pdfkeywords={CMS, flow harmonic, charm quark, cumulants, flow fluctuations}}

\maketitle

A strongly coupled quark-gluon plasma (QGP), has 
been studied in nucleus-nucleus collisions at the BNL RHIC~\cite{Arsene:2004fa,Back:2004je,Adams:2005dq,Adcox:2004mh} 
and CERN LHC~\cite{Muller:2012zq,Armesto:2015ioy}.
This medium exhibits the behavior of a nearly-perfect liquid~\cite{Heinz:2013th,Gale:2013da}. 
The azimuthal anisotropy of produced hadrons, resulting from pressure-driven expansion, is a powerful tool to study QGP dynamics and can be characterized by the Fourier coefficients ({\vn}) of the hadrons' azimuthal angle ($\phi$) distribution~\cite{Poskanzer:1998yz}. 
The second-order Fourier coefficient ({\vtwo}), known as elliptic flow, of low transverse momentum ({\pt}) particles reflects 
the QGP response to the average initial collision geometry and its event-by-event fluctuations~\cite{Poskanzer:1998yz}. 
The \vtwo coefficient is also influenced by the path length dependence 
of parton energy loss at high {\pt}~\cite{Ollitrault:2009ie,Katz:2019fkc,Katz:2019qwv}.

Charm and beauty (heavy-flavor) quarks are produced in the initial stages
of a collision via hard scattering processes~\cite{Harris:2015hra}. 
At the LHC, a significant elliptic flow is observed for mesons containing a
charm quark, namely prompt {\PDz}~\cite{Acharya:2017qps,Sirunyan:2017plt,ALICE:2018gif,ALICE:2020iug,Sirunyan:2020dop}
and {\PJGy}~\cite{Khachatryan:2016ypw,Acharya:2017tgv,ALICE:2020pvw}, and 
for leptons from heavy-flavor hadron decays~\cite{ATLAS:2020yxw,ALICE:2020hdw}. However, the first
measurements with mesons containing unambiguous beauty quarks, specifically the \PgUa and {\PgUb}~\cite{Acharya:2019hlv,CMS:2020efs},
show \vtwo values compatible with zero. The \PDz meson \vtwo has been measured
using a two-particle cumulant method $\vtwotwo$~\cite{Luzum:2012da}, 
at RHIC~\cite{Adler:2002pu,Adamczyk:2017xur} and LHC~\cite{Acharya:2017qps,Sirunyan:2017plt,ALICE:2018gif,ALICE:2020iug,Sirunyan:2020dop}.
This method correlates a \PDz meson with each charged particle in the event.
The results indicate that low-{\pt} charm quarks
are strongly coupled to the QGP,
as reproduced by hydrodynamic models~\cite{Gale:2013da}. 

The magnitude of event-by-event fluctuations~\cite{Voloshin:2007pc} of azimuthal anisotropy harmonics from
heavy-flavor quarks has not been experimentally measured.
Multiparticle correlation techniques involving four or more particles, $\vtwo\{n\}$, with $n\geq4$~\cite{Bilandzic:2010jr}, 
allow direct access to cumulants of the \vtwo probability density distribution.
The technique has been widely applied in the light-flavor sector to extract the magnitude of \vtwo fluctuations, 
which is then used to constrain fluctuations of the initial-state geometry. It has been recently proposed that
for hard probes (such as high-{\pt} jets, and heavy-flavor hadrons), fluctuations of anisotropy harmonics are
not only influenced by the initial-state geometry, but are also sensitive to final-state fluctuations of
energy loss when these hard probes propagate in the QGP medium~\cite{Betz:2016ayq}.
Therefore, measurements of \vtwofour and its ratio to \vtwotwo for heavy-flavor hadrons have the potential to set constraints 
on the mechanism of heavy-quark energy loss, especially how it fluctuates on an event-by-event basis in QGP. 

In this letter, the prompt \PDz meson \vtwo coefficient is measured for
the first time using four-particle correlations, and the ratio \vtworatio is presented. 
These measurements use data from lead-lead ({\PbPb}) collisions 
at a nucleon-nucleon center-of-mass energy of $\sqrtsNN=5.02\TeV$, collected by the CMS detector at the LHC in 2018. 
The behavior of \vtwo is examined in the rapidity ($y$) range $\abs{y}<1$ 
over the \pt range of 2--15\GeV, and in the event centrality classes 
(\ie the percentage ranges of the total inelastic hadronic cross section) 
of 10--30\% and 30--50\%. A 0\% centrality corresponds to the largest overlap of the two nuclei.
The centrality dependence of \vtwo is also measured
over the broader range of 5--60\% for $2<\pt<8\GeV$.
Tabulated results are provided in the HEPData record for this analysis~\cite{hepdata}.

The CMS apparatus~\cite{Chatrchyan:2008zzk} is a multipurpose, nearly hermetic detector,
designed to trigger on~\cite{Sirunyan:2020zal,Khachatryan:2016bia} and identify electrons,
muons, photons, and hadrons~\cite{CMS:2015xaf,CMS:2018rym,CMS:2015myp,Chatrchyan:2014fea}.
In this analysis, the information from two subdetectors were used: the silicon inner tracker, 
which measures charged particles within the range of pseudorapidity $\abs{\eta}<3.0$; 
and the hadronic forward (HF) calorimeters, made of steel and quartz fibres, 
which extend the pseudorapidity coverage provided by the barrel and endcap detectors to about $\abs{\eta}<5.0$, 
and are segmented to form $0.175{\times}0.175$ ($\Delta\eta{\times}\Delta\phi$) towers. 

The data analyzed consist of 4.27\ten{9} minimum bias events, 
corresponding to an integrated luminosity of 0.58\nbinv. 
The events are triggered by requiring signals above thresholds in the range of ${\sim}$6--12\GeV in both sides of
the HF calorimeters~\cite{Khachatryan:2016bia}. 
Events must also have at least one reconstructed primary vertex within 15\cm of the interaction point along the beam axis. 
The primary vertex is selected as the one with the highest track multiplicity in the event. 
The effects from concurrent interactions in the same bunch crossing were shown to be negligible. 
The centrality is calculated using the HF calorimeters~\cite{CMS:2011iwn}.

Monte Carlo (MC) event samples are simulated containing either prompt or nonprompt \PDz mesons; 
the latter originate from beauty hadron decays.
The simulated events are generated using \PYTHIA 8.212~\cite{Sjostrand:2014zea}, tune
CP5~\cite{Sirunyan:2019dfx}, and embedded into MC \PbPb events from \HYDJET 1.9~\cite{Lokhtin:2005px}.
The prompt \PDz meson event sample is
employed to define signal selections and efficiency corrections, 
while the other sample is used to estimate systematic uncertainties from nonprompt \PDz contamination.

Both \PDz and \PaDz mesons are reconstructed via the process $\PDz\to\PGpp+\PKm$ ($\PaDz\to\PGpm+\PKp$), 
with a branching fraction of ($3.95\pm0.03$)\%~\cite{PDG2020}.
This is accomplished by combining pairs of oppositely charged tracks having an 
invariant mass (\minv) within ${\pm}200\MeV$ of the world-average \PDz mass of 
1865\MeV~\cite{PDG2020}. Tracks are required to have $\pt>1.0\GeV$ and $\abs{\eta}<2.4$
and must satisfy high-purity quality criteria~\cite{Chatrchyan:2014fea}. 
Two \PDz candidates for each pair of selected tracks are considered by assuming one track has the pion mass, while the other has the kaon mass, and vice versa. Kinematic fits~\cite{Forden:1985xa} are performed to reconstruct the
decay (secondary) vertex of each \PDz candidate. A boosted
decision tree (BDT) algorithm, as implemented in
the $\textsc{tmva}$ software package~\cite{Hocker:2007ht}, maximizes 
the statistical significance of prompt \PDz meson signals.
Particle pairs having the same charge, and again assumed to be a pion and kaon, are used 
as the background distribution for training the BDT.
This analysis uses the same BDT parameters as Ref.~\cite{Sirunyan:2020dop}.

\begin{figure}[htb]
\centering
\includegraphics[width=0.85\linewidth]{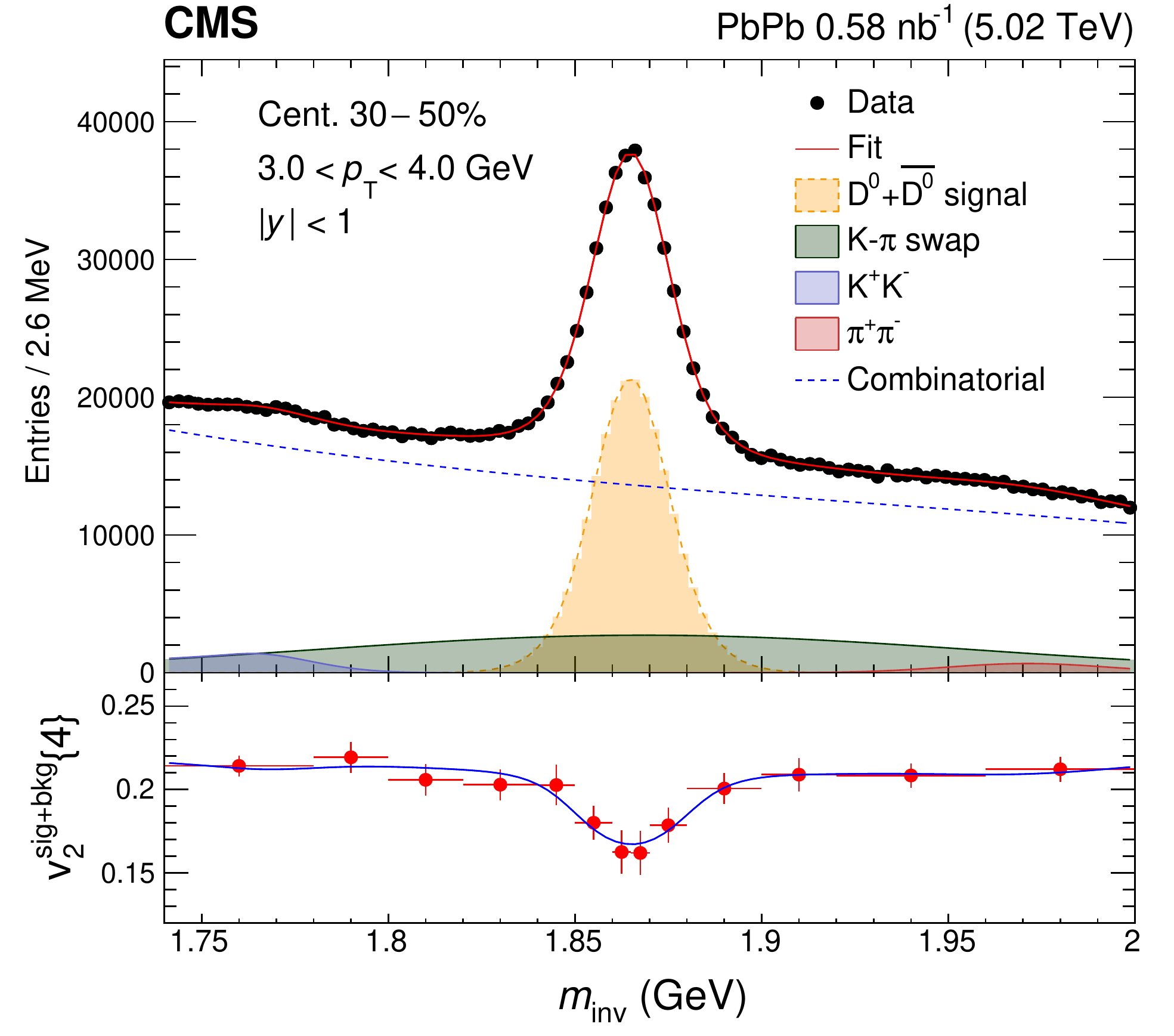}
\caption{An example of the two-step fit of the mass spectrum (upper) and \vtwosigbkg (lower)
        in the \pt interval 3--4\GeV for the centrality class 30--50\%.}
\label{fig:2stepsfit}
\end{figure}

This analysis shares the same data sets 
and uses a similar procedure to that described in Ref.~\cite{Sirunyan:2020dop}, in which the \PDz meson \vtwo is measured 
using the two-particle correlation (or cumulant) method, $\vtwotwo({\PDz})$, where the \PDz meson 
\vtwo signal is extracted by correlating a \PDz meson with reference particles measured in the HF detectors. 
To measure the differential second-order (elliptic) harmonic from 
the four-particle cumulant, $\vtwofour({\PDz})$~\cite{Bilandzic:2010jr}, a first step involves either two- or four-particle correlations calculated using energy deposits in the HF towers to obtain elliptic harmonics of reference particles. 
Here, each HF tower is used to represent one or more particles with a weight applied corresponding to its deposited transverse 
energy in the calculation of cumulants when averaging over all HF towers, as detailed below.
The two- and four-particle azimuthal correlations for the $n$th harmonic are defined as
\begin{linenomath}
\begin{equation}
  \llangle {2} \rrangle = \llangle \re^{in(\phi_{1}^{\textit{a}}-\phi_{2}^{\textit{b}})} \rrangle,\quad
  \llangle {4} \rrangle = \llangle \re^{in(\phi_{1}^{\textit{a}}+\phi_{2}^{\textit{a}}-\phi_{3}^{\textit{b}}-\phi_{4}^{\textit{b}})} \rrangle.
  \label{eqn:azimuthal_corr_ref_2P_4P}
\end{equation}
\end{linenomath}
Here, $\phi_{\mathrm{j}}$ (j = 1, \ldots, 4) are the azimuthal angles of one unique
combination of multiple particles in an event and the double average
symbol $\llangle\cdots\rrangle$ indicates that the average is taken over
all unique particle combinations and for all events.
In addition, the superscripts $\textit{a}$ and $\textit{b}$ indicate towers
chosen from two different HF calorimeters, $\mathrm{HF-}$ ($-5<\eta<-3$) or $\mathrm{HF+}$ ($3<\eta<5$). 
In a second step, the four-particle cumulant of reference particle azimuthal correlations, \cnfour, is calculated as~\cite{Bilandzic:2013kga,Khachatryan:2015waa,Borghini:2001vi,Bilandzic:2010jr}
\begin{linenomath}
\begin{equation}
   \cnfour = \llangle {4} \rrangle - 2\, {\llangle2\rrangle}^2.
\label{eqn:cumuant_ref}
\end{equation}
\end{linenomath}
To measure the prompt \PDz meson \vtwo coefficient, the $\phi_{1}^{\textit{a}}$ from an HF tower in
Eq.(\ref{eqn:azimuthal_corr_ref_2P_4P}) is replaced with a \PDz 
candidate's azimuthal angle selected within the tracker
acceptance $\abs{\eta}<2.4$.
To suppress the nonflow effects from sources such as resonance decays or jets, 
in the two-particle cumulant method, a tower with $\phi_2$ is selected from the HF calorimeter ($\mathrm{HF+}$ or $\mathrm{HF-}$) having the opposite 
$\eta$ sign as that of the \PDz candidate. For the four-particle correlations method, 
$\phi_{2}^{\textit{a}}$ is picked from the HF detector having the same $\eta$ sign as the \PDz candidate, 
but $\phi_{3}^{\textit{b}}$ and $\phi_{4}^{\textit{b}}$ are chosen from the other HF detector. 
Studies performed with simulated events indicate that nonflow effects 
are negligible when measuring \vtwotwo with these $\eta$ gaps~\cite{Khachatryan:2015waa,Chatrchyan:2013nka}. 
These modified particle correlators involving a \PDz meson are denoted by
$\llangle{2'}\rrangle$ and $\llangle{4'}\rrangle$. The differential four-particle
cumulant of \PDz mesons is then defined as~\cite{Borghini:2001vi,Bilandzic:2010jr},
\begin{linenomath}
\begin{equation}
   \dnfour = \llangle{4'}\rrangle - 2 \, {\llangle{2'}\rrangle}{\llangle2\rrangle}.
\label{eqn:cumulant_POI}
\end{equation}
\end{linenomath}
Finally, with respect to the reference four-particle cumulants,
the differential four-particle $\vn({\PDz})$ coefficients are extracted as in Refs.~\cite{Borghini:2001vi,Bilandzic:2010jr},
\begin{linenomath}
\begin{equation}
   \vnfour({\PDz}) = -\frac{\dnfour}{(-\cnfour)^{3/4}}, 
\label{eqn:diff_flow_4}
\end{equation}
\end{linenomath}
which includes contributions of both true signal and background \PDz candidates.
To separate the \vtwo signal of \PDz mesons 
($\vtwo^{\text{sig}}\{4\}$)
from background candidates 
($\vtwo^{\text{bkg}}\{4\}$),
the same two-step fitting procedure as in Ref.~\cite{Sirunyan:2020dop} is performed. 
First, the invariant mass spectrum of all \PDz 
candidates is fit using a formula containing five components:
(i) A sum of two Gaussian functions having the same mean but different widths are used
for the \PDz signal; (ii) a single Gaussian
function describes the invariant mass spectrum of
\PDz candidates with an incorrect mass assignment resulting from the exchange
of the kaon and pion designations; (iii) a Crystal Ball function~\cite{CrystalBallRef} 
is used for the processes $\PDz\to\PKp\PKm$~\cite{Sirunyan:2020obi}; (iv)
another Crystal Ball function to describe $\PDz\to\Pgpp\Pgpm$~\cite{Sirunyan:2020obi};
(v) a third-order polynomial is used to model the combinatorial background.
The first four components are initialized by values
calculated using simulated events, and their widths are allowed to vary
with a common scale factor during the fit to data.
Using the signal and background \PDz candidate yield fraction extracted 
from the invariant mass fit, the measured \vtwo data of all \PDz candidates, \vtwosigbkg
can then be decomposed into the \vtwo values of signal and background 
\PDz candidates, by fitting to a linear combination of the two components.
An example of the full fitting procedure is shown in Fig.~\ref{fig:2stepsfit}.
The influence from the \PDz meson \vtwo signal can be clearly seen in the lower panel as a dip in the \vtwosigbkg distribution.

Statistical uncertainties are evaluated from data with the method described in
Refs.~\cite{Sirunyan:2017pan,Khachatryan:2015waa}. 
The data are divided into 20 equal subsets, and the 
standard deviation of the resulting cumulant distribution is
used to estimate the statistical uncertainty.

The sources of systematic uncertainties include the \PDz meson
BDT selection (\ie, the choice of the working point), the background invariant mass probability
distribution (PD), the PD of the \vtwo background, 
the detector acceptance and \PDz meson reconstruction efficiency correction of the \PDz meson yield, as well as nonprompt \PDz meson contamination. 
The uncertainties in the \vtworatio ratios account for the correlations between uncertainty sources for \vtwofour and \vtwotwo. 
The systematic uncertainty of the BDT selection is assigned by varying up and down the BDT discriminant requirement. 
The magnitudes of these variations depend on the collision centrality and are derived by comparing 
the BDT discriminant requirement optimization in simulation and in a subset of data events. 
It is 0.002--0.004 for \vtwofour 
and 0.020--0.035 for \vtworatio. The systematic
uncertainties from the mass background PD are evaluated by changing
the default third-order polynomial function to a second-order
polynomial or exponential function, and are between 0.002--0.005 for \vtwofour
and 0.004--0.019 for \vtworatio. The systematic uncertainties
from the \vtwo background PD are evaluated by changing the default
linear function to a second-order polynomial or a constant function,
and are 0.002--0.005 for \vtwofour and 0.003--0.014
for \vtworatio. 
Although the efficiency of selecting \PDz mesons essentially cancels when measuring the \vtwo, 
the systematic uncertainty from the efficiency correction
is evaluated by comparing results with and without applying efficiency corrections to the \PDz meson yield.
The \PDz yield corrections are applied in intervals
of \pt for $\abs{y}<1$, using the acceptance and efficiency values obtained
from simulated events. This correction yields the uncertainties of 0.004--0.016 for \vtwofour and 0.033--0.116 for \vtworatio for the $2<\pt<3\GeV$ bin (with the ranges corresponding to the variation between the centrality bins of 10--30\% and 30--50\%), and becomes negligible at higher \pt values. 
The uncertainties from efficiency correction are also quoted in the {\pt}-integrated ($2<\pt<8\GeV$) \vtwo results in different centralities in the range of 5--60\%, with an average value of 0.006 for \vtwofour and of 0.015 for \vtworatio. 
The systematic uncertainties
from the nonprompt \PDz contamination (2--5\%) are evaluated by
using the relative uncertainty estimated in Ref.~\cite{Sirunyan:2020dop} for \vtwotwo,
and are 0.001--0.005 for \vtwofour. All the different sources are added together 
in quadrature and the total uncertainty is 0.008--0.018 for \vtwofour
and 0.021--0.121 for \vtworatio. 

\begin{figure*}[htb]
\centering
\includegraphics[width=0.90\linewidth]{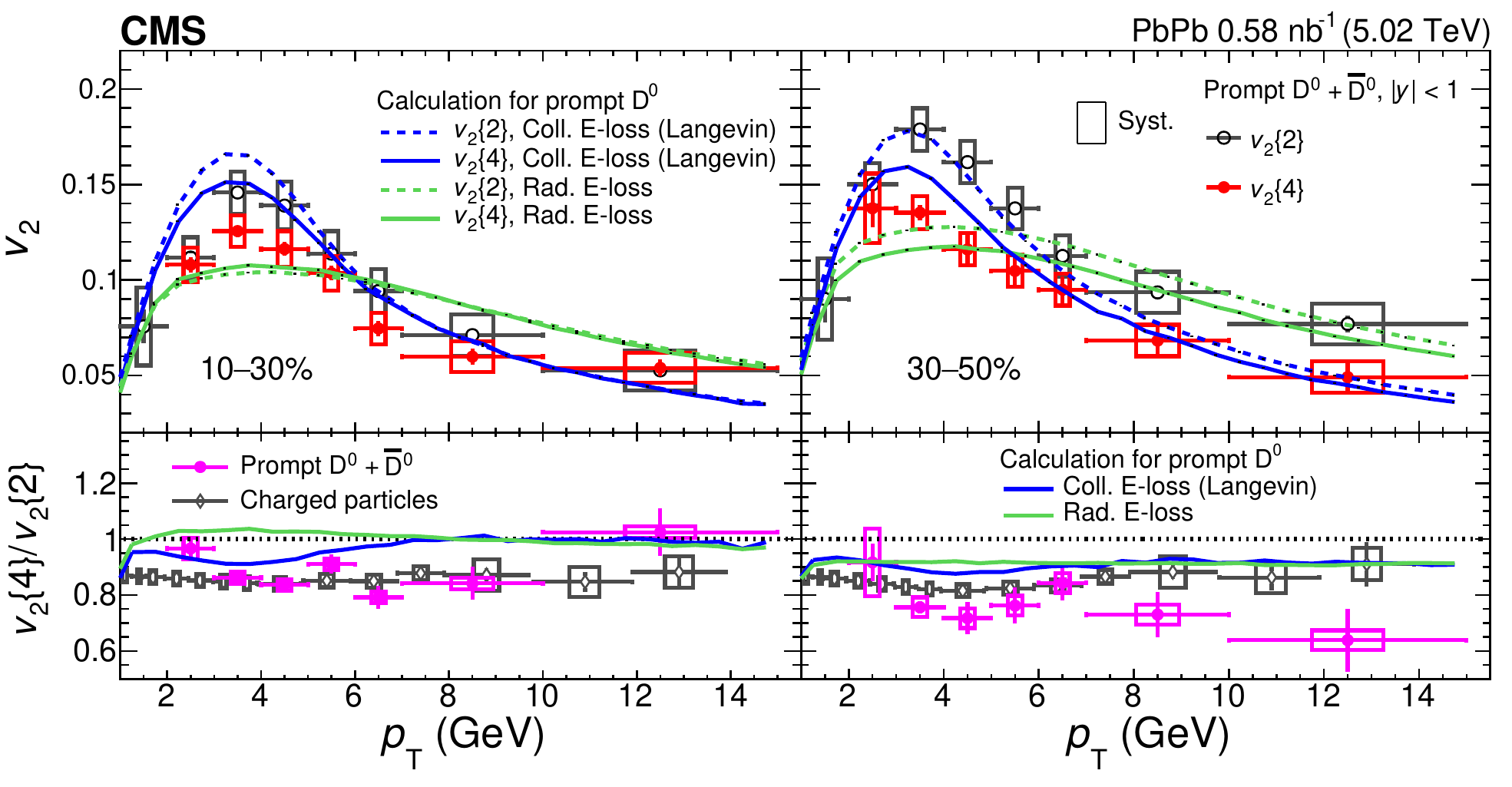}
\caption{
  Upper panel: prompt \PDz meson \vtwotwo and \vtwofour
  coefficients as a function of \pt, for the centrality classes 10--30\% (left)
  and 30--50\% (right). 
  The lines indicate calculations from the DABMod model~\cite{Katz:2019fkc,Katz:2019qwv}, 
  with solid (dashed) lines indicating \vtwofour ({\vtwotwo}) values. 
  Blue lines include Langevin dynamics and green lines include radiative energy loss (E-loss). 
  Lower panel: the prompt \PDz meson \vtworatio ratios are shown and
  compared to those for charged particles in the pseudorapidity range $\abs{\eta}<1$~\cite{Sirunyan:2017pan}.
  The vertical bars represent statistical uncertainties and the open boxes
  denote the systematic uncertainties.
}
\label{fig:v24_vs_pT}
\end{figure*}

Figure~\ref{fig:v24_vs_pT} shows \vtwofour results of prompt \PDz
mesons (upper panel) within the mid-rapidity range $\abs{y}<1$ as a function of \pt.
These \vtwo values are measured in the centrality classes 10--30\% and 30--50\%. 
The \vtwotwo values, measured previously by CMS in Ref.~\cite{Sirunyan:2020dop}, are also shown for comparison.
As previously observed for \vtwotwo, the measured \vtwofour
values rise with increasing \pt, up to a maximum near $\pt\approx3.5\GeV$,
and then diminish.
The \vtwofour values are below the \vtwotwo measurements, 
with the difference being more pronounced above 3\GeV and for the 30--50\% centrality range.
A similar observation has been found for all charged particles in the event
~\cite{Sirunyan:2017pan}, which is predicted by initial-state geometry fluctuations modeled by
using Bessel--Gaussian and elliptic power eccentricity distributions
~\cite{Voloshin:2007pc,Sirunyan:2017fts}. The elliptic power distribution is a two-parameter distribution, 
where one of the parameters corresponds to the intrinsic eccentricity, 
while the other parameter controls the magnitude of eccentricity fluctuations.
Theoretical calculations for prompt \PDz meson \vtwo based on
a state-of-the-art \PD and \PB meson modular simulation
code (called DABMod~\cite{Katz:2019fkc,Katz:2019qwv}) 
with the option of turning on energy loss by gluon radiation or 
alternatively by elastic collisions described by Langevin dynamics
during the heavy-quark propagation, are also shown in the upper
panel of Fig.~\ref{fig:v24_vs_pT}. The radiative energy loss process is expected to
be the dominant phenomenon in the high-{\pt} region. Langevin
dynamics, which describe the propagation of heavy quarks in the medium as a Brownian motion, can account
for collisional processes using Langevin-like equations~\cite{Katz:2019fkc} in the low- and intermediate-{\pt} region. 
Both models seem to capture the general trends of the data,
without reproducing them quantitatively.

To further investigate the underlying physics processes behind
elliptic flow fluctuations of charm quarks, the ratios \vtworatio are
presented as a function of \pt, up to 15\GeV, in the lower panel of
Fig.~\ref{fig:v24_vs_pT}. Generally speaking, a larger deviation of \vtworatio 
ratios from unity indicates a larger magnitude of flow fluctuations.
The same ratios for charged particles (dominated by light-flavor hadrons) are shown. 
The ratios for prompt \PDz mesons are consistent with those for charged particles.
The roughly flat behavior of the ratios at low \pt suggests that 
initial-state geometry fluctuations are likely the dominant source of flow fluctuations there~\cite{Katz:2019fkc}.
The ratios based on the DABMod model for \PDz mesons~\cite{Katz:2019fkc,Katz:2019qwv}, also shown in Fig.~\ref{fig:v24_vs_pT} (bottom),
lie systematically above the data, suggesting an underestimation of the magnitude of flow fluctuations in the data.

\begin{figure}[htb]
\centering
\includegraphics[width=0.85\linewidth]{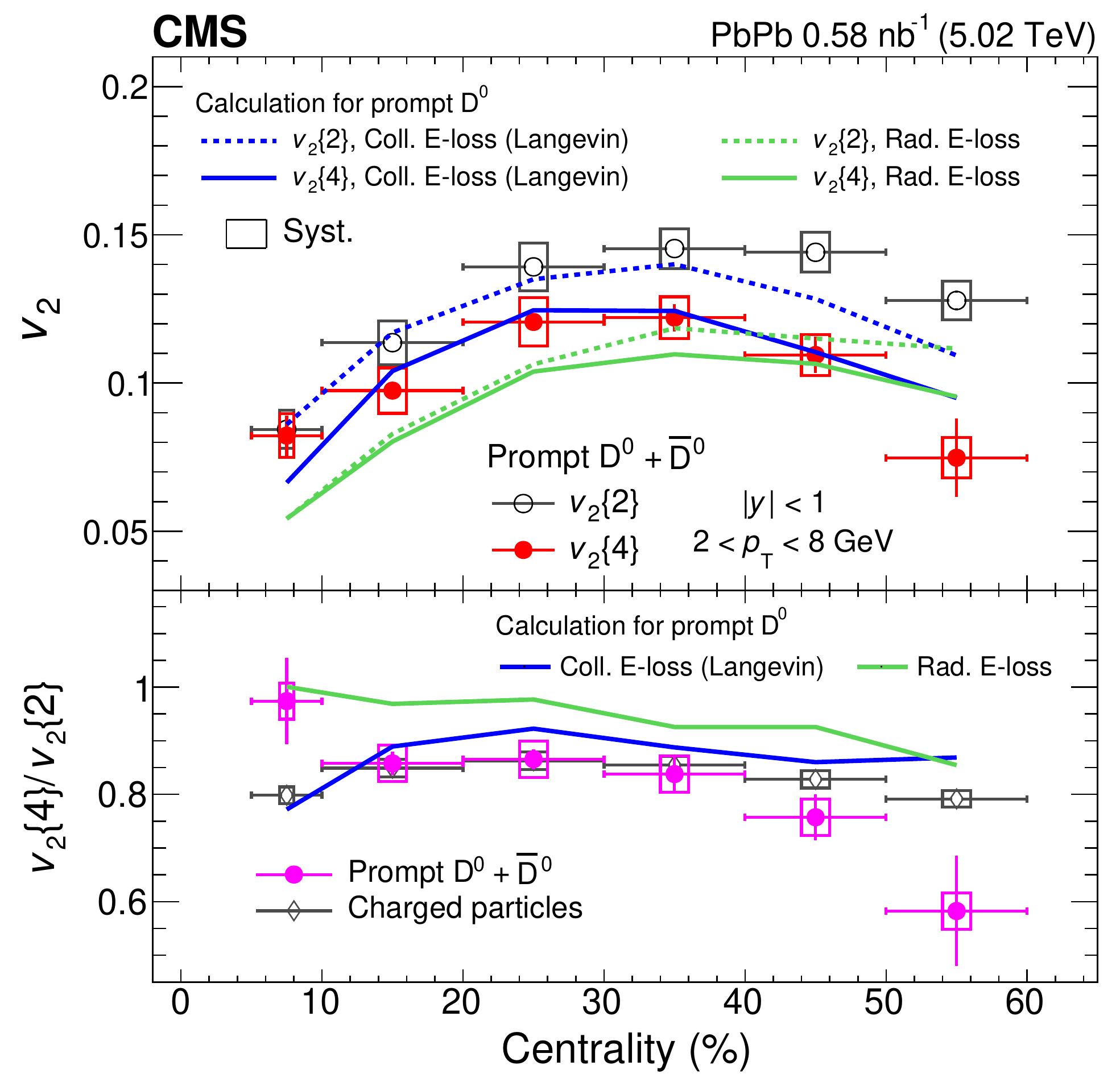}
\caption{
  Upper panel: the prompt \PDz meson \vtwotwo and \vtwofour as a function of centrality. 
  The lines indicate calculations from the DABMod model~\cite{Katz:2019fkc,Katz:2019qwv},
  with solid (dashed) lines indicating \vtwofour ({\vtwotwo}) values.
  Blue lines include Langevin dynamics and green lines include radiative energy loss (E-loss). 
  Lower panel: the prompt \PDz meson \vtworatio are
  compared to the same ratio for charged particles in the pseudorapidity range $\abs{\eta}<1$~\cite{Sirunyan:2017pan}.
  The vertical bars represent statistical uncertainties and open boxes
  denote the systematic uncertainties. 
}
\label{fig:v24_vs_cent}
\end{figure}

The {\pt}-integrated results of \vtwofour for $2<\pt<8\GeV$ and
$\abs{y}<1$ are shown as a function of centrality from 5 to 60\% in
Fig.~\ref{fig:v24_vs_cent}. The \vtwotwo values measured previously by CMS in
Ref.~\cite{Sirunyan:2020dop} are plotted for
comparison. The \vtworatio ratios are shown in the lower panel of
Fig.~\ref{fig:v24_vs_cent}.
The prompt \PDz data are also compared to those of inclusive charged
particles within the range $\abs{\eta}<1$ and $2<\pt<8\GeV$. 

Similar to the \PDz meson \vtwotwo coefficient, the \PDz meson 
\vtwofour value increases with centrality in the 5--40\% range, 
and then decreases for more peripheral collisions. This trend is qualitatively
reproduced by calculations incorporating an interplay of initial-state geometry 
and parton energy loss in QGP. Within the 10--40\% centrality range,
the \vtworatio ratios are almost identical between
prompt \PDz mesons and inclusive charged particles within uncertainties. 
This indicates that, within this centrality range, the dominant source of flow fluctuations 
for heavy-flavor is similar to that for soft light-flavor particles, namely initial-state geometry fluctuations, 
and therefore the contribution from final-state fluctuations is small. 
The hint of different trends in \vtworatio between \PDz mesons and charged
particles seen in the most central and most peripheral events could indicate that
fluctuations from final-state effects, such as parton energy loss, in hard processes  
become visible for charm mesons~\cite{Katz:2019fkc}. For example, as the system size
becomes smaller for peripheral events, the number of scatterings a hard probe experiences with QGP 
will decrease, leading to larger fluctuations in the energy loss on an event-by-event basis.
However, the experimental uncertainties
are still large, with the difference of ${\sim}2$ standard deviations between the values.
Calculations based on the DABMod model~\cite{Katz:2019fkc,Katz:2019qwv} assuming 
collisional (or Langevin dynamics) and radiative energy loss processes are also shown in Fig.~\ref{fig:v24_vs_cent}. 
A better description of the experimental data is obtained using the Langevin dynamics,
although no increase or decrease for the most central or peripheral events, respectively, is predicted.

In summary, the first measurements of the elliptic flow for prompt \PDz and \PADz mesons using 
a four-particle cumulant method are presented.
These \vtwofour values are systematically lower than
the measured two-particle elliptic flow values, \vtwotwo, indicating the presence of 
event-by-event fluctuations in the flow signal~\cite{Voloshin:2007pc}.
To further investigate the origin of \vtwo fluctuations, 
\vtworatio ratios of prompt \PDz mesons are compared to those of
light-flavor hadrons. Similar trends for both charm mesons and light-flavor
hadrons are observed, suggesting that the dominant contribution to \vtwo fluctuations 
comes from the initial geometry. 
An indication of splitting of the \vtworatio ratios between charm mesons
and light-flavor hadrons in the most central and
most peripheral events is seen, which may suggest an additional contribution, such as energy loss fluctuations.
Model calculations implementing collisional energy loss mechanisms provide a better description of the data than those considering radiative energy loss.

\begin{acknowledgments}
  We congratulate our colleagues in the CERN accelerator departments for the excellent performance of the LHC and thank the technical and administrative staffs at CERN and at other CMS institutes for their contributions to the success of the CMS effort. In addition, we gratefully acknowledge the computing centers and personnel of the Worldwide LHC Computing Grid and other centers for delivering so effectively the computing infrastructure essential to our analyses. Finally, we acknowledge the enduring support for the construction and operation of the LHC, the CMS detector, and the supporting computing infrastructure provided by the following funding agencies: BMBWF and FWF (Austria); FNRS and FWO (Belgium); CNPq, CAPES, FAPERJ, FAPERGS, and FAPESP (Brazil); MES and BNSF (Bulgaria); CERN; CAS, MoST, and NSFC (China); MINCIENCIAS (Colombia); MSES and CSF (Croatia); RIF (Cyprus); SENESCYT (Ecuador); MoER, ERC PUT and ERDF (Estonia); Academy of Finland, MEC, and HIP (Finland); CEA and CNRS/IN2P3 (France); BMBF, DFG, and HGF (Germany); GSRI (Greece); NKFIA (Hungary); DAE and DST (India); IPM (Iran); SFI (Ireland); INFN (Italy); MSIP and NRF (Republic of Korea); MES (Latvia); LAS (Lithuania); MOE and UM (Malaysia); BUAP, CINVESTAV, CONACYT, LNS, SEP, and UASLP-FAI (Mexico); MOS (Montenegro); MBIE (New Zealand); PAEC (Pakistan); MSHE and NSC (Poland); FCT (Portugal); JINR (Dubna); MON, RosAtom, RAS, RFBR, and NRC KI (Russia); MESTD (Serbia); MCIN/AEI and PCTI (Spain); MOSTR (Sri Lanka); Swiss Funding Agencies (Switzerland); MST (Taipei); ThEPCenter, IPST, STAR, and NSTDA (Thailand); TUBITAK and TAEK (Turkey); NASU (Ukraine); STFC (United Kingdom); DOE and NSF (USA).

\end{acknowledgments}
\bibliography{auto_generated}
\cleardoublepage \appendix\section{The CMS Collaboration \label{app:collab}}\begin{sloppypar}\hyphenpenalty=5000\widowpenalty=500\clubpenalty=5000\cmsinstitute{Yerevan~Physics~Institute, Yerevan, Armenia}
A.~Tumasyan
\cmsinstitute{Institut~f\"{u}r~Hochenergiephysik, Vienna, Austria}
W.~Adam\cmsorcid{0000-0001-9099-4341}, J.W.~Andrejkovic, T.~Bergauer\cmsorcid{0000-0002-5786-0293}, S.~Chatterjee\cmsorcid{0000-0003-2660-0349}, M.~Dragicevic\cmsorcid{0000-0003-1967-6783}, A.~Escalante~Del~Valle\cmsorcid{0000-0002-9702-6359}, R.~Fr\"{u}hwirth\cmsAuthorMark{1}, M.~Jeitler\cmsAuthorMark{1}\cmsorcid{0000-0002-5141-9560}, N.~Krammer, L.~Lechner\cmsorcid{0000-0002-3065-1141}, D.~Liko, I.~Mikulec, P.~Paulitsch, F.M.~Pitters, J.~Schieck\cmsAuthorMark{1}\cmsorcid{0000-0002-1058-8093}, R.~Sch\"{o}fbeck\cmsorcid{0000-0002-2332-8784}, M.~Spanring\cmsorcid{0000-0001-6328-7887}, S.~Templ\cmsorcid{0000-0003-3137-5692}, W.~Waltenberger\cmsorcid{0000-0002-6215-7228}, C.-E.~Wulz\cmsAuthorMark{1}\cmsorcid{0000-0001-9226-5812}
% ...
\vskip\cmsinstskip
% ...
1:~Also at TU Wien, Wien, Austria\\
% ...
\end{sloppypar}
\end{document}